\documentclass{article} 
\usepackage{latexsym, amsmath, amssymb, enumerate} 
\newtheorem{theorem}{Theorem} 
\newtheorem{proposition}{Proposition} 
\newtheorem{lemma}{Lemma}

\numberwithin{equation}{section} 
\newcommand{\supp}{\mbox{ supp }}
\newcommand{\g}{g^\sharp}
\newcommand{\h}{h^\sharp}
\newcommand{\farg}{(t-c^{-1}|y-x|, y, p)} 
\newcommand{\intL}{\int_{|y-x| \leq ct}} 
\newcommand{\intG}{\int_{|y-x| > ct}}
\newcommand{\intE}{\int_{|y-x| = ct}} 
\newcommand{\intw}{\int_{|\omega|=1}} 
\newcommand{\sqcp}{\sqrt{1 + c^{-2}p^2}} 
\newcommand{\cp}{1 + c^{-2}p^2} 
\newcommand{\op}{1 + c^{-1} \omega \cdot \widehat{p}} 
 
\newcommand{\owp}{\omega\wedge\widehat{p}} 
\newcommand{\oap}{\omega + c^{-1} \widehat{p}} 
\newcommand{\wdp}{\omega \cdot \widehat{p}} 
\newcommand{\prfe}{\hfill $\Box$ 

						  \smallskip}

\title{The non-relativistic limit \\ of the Nordstr\"om-Vlasov system} 
\date{} 
\author{SIMONE CALOGERO\footnote{E-Mail: mg026@math.chalmers.se}\\
\textit{Department of Mathematics, Chalmers University} \\ \textit{S-412 96 G\"oteborg, Sweden}\\[3mm]
HAYOUNG LEE\footnote{E-mail: hayoung@aei.mpg.de}\\
\textit{Albert Einstein Institute, Am M\"uhlenberg 1}\\ \textit{D-14476 Golm (bei Potsdam), Germany}}

\begin{document} 
\maketitle
\begin{abstract}
The Nordstr\"om-Vlasov system provides an interesting relativistic generalization of the Vlasov-Poisson
system in the gravitational case, even though there is no direct physical application.
The study of this model will probably lead to a better mathematical understanding of the class of
non-linear systems consisting of hyperbolic and transport equations. 
In this paper it is shown that solutions of the Nordstr\"om-Vlasov system
converge to solutions of the Vlasov-Poisson system in a pointwise sense as the speed of light tends to infinity, 
providing a further and rigorous justification of this model as a \textit{genuine} relativistic generalization of the Vlasov-Poisson system. 
\end{abstract}
\pagestyle{myheadings}
\thispagestyle{plain}
\markboth{SIMONE CALOGERO AND HAYOUNG LEE}{THE NORDSTR\"OM-VLASOV SYSTEM}
 
\section{Introduction}

Kinetic models of collisionless matter have many important physical applications. In astrophysics, for example, the stars of a
galaxy are often modelled as a large ensemble of particles in which collisions are sufficiently rare to be neglected. 
The distribution $f_\infty$ of particles in the phase-space satisfies the Vlasov-Poisson system:
\begin{gather}
\partial_tf_\infty+p\cdot\nabla_xf_\infty -\nabla_xU\cdot\nabla_pf_\infty=0,\label{vlasovpoisson1}\\
\Delta_x U=4\pi\gamma\rho_\infty,\quad\gamma=1,\quad\rho_\infty=\int_{\mathbb{R}^3}f_\infty\,dp.\label{vlasovpoisson2}
\end{gather}
In the previous equations, $f_\infty=f_\infty(t,x,p)$ gives the probability density to find a particle (star) at time $t$ at position
$x$ with momentum $p$, where $t\in\mathbb{R},\,x\in\mathbb{R}^3,\,p\in\mathbb{R}^3$. $U=U(t,x)$ is the mean Newtonian potential 
generated by the stars. 

By replacing $\gamma=-1$ in (\ref{vlasovpoisson2}) one obtains the Vlasov-Poisson
system in the plasma physics case. Here the particles are charges and $U$ is the electrostatic potential which they create collectively. 
We consider a single species of particle in both cases. 
The applications of these Vlasov-Poisson systems are restricted to the situations 
where the relativistic effects are negligible, i.e., low velocities and weak fields. 
Otherwise the dynamics has to be described by the
relativistic Vlasov-Maxwell system in plasma physics and by the Einstein-Vlasov system in stellar dynamics.

The two Vlasov-Poisson models are very similar to each other 
and no substantial difference arises in the question of global existence of classical solutions, which is by now well-understood (cf. \cite{LP,Pf,R,Sch2}). 
As opposed to this, the relativistic models have very different structure and so far they have been considered
separately. In the gravitational case, global existence of (asymptotically flat) solutions for the Einstein-Vlasov system 
is known only for small data with spherical symmetry \cite{RR1}. 
For the relativistic Vlasov-Maxwell system the theory is more developed,
cf. \cite{C2,DL}, \cite{GSh1}--\cite{KS}, \cite{R2}. 
However global existence and uniqueness of classical solutions 
for large data in three dimensions is still open. 

In a recent paper \cite{C}, a different relativistic generalization to the Vlasov-Poisson system in the stellar dynamics case has been 
considered,
in which the Vlasov dynamics is coupled to a relativistic {\it scalar} theory of gravity which goes back, essentially, to Nordstr\"om
\cite{No}. More precisely, the gravitational theory considered in \cite{C} corresponds to a reformulation of Nordstr\"om's theory due to 
Einstein and Fokker (see \cite{EF}). The resulting system has been called Nordstr\"om-Vlasov system and reads 
\begin{align} 
&-\partial_t^2\phi + c^2\Delta_x\phi  
			= 4\pi\int\frac{f \, dp}{\sqcp}, \label{NV1}\\ 
&\partial_tf + \widehat{p}\cdot\nabla_xf  
       - \big[S(\phi)p+\frac{c^2\nabla_x\phi}{\sqcp}\big]\cdot\nabla_pf 
 = 4 S(\phi)f, \label{NV2} 
\end{align} 
where 
$$ 
p^2 = |p|^2,\quad  
\widehat{p} = (1+c^{-2}p^2)^{-1/2}p,\quad  
S = \partial_t+\widehat{p}\cdot\nabla_x. 
$$
Here $f = f(t, x, p)$, $\phi = \phi(t, x)$ and
$c$ denotes the speed of light. A solution $(f,\phi)$ of this system  
is interpreted as follows. The spacetime is a Lorentzian manifold with a conformally flat metric which, in the coordinates  
$(ct,x)$, takes the form 
$$
g_{\mu\nu}=e^{2\phi} \textrm{diag}(-1,1,1,1). 
$$
Throughout the paper Greek indices $\mu$, $\nu$ and $\sigma$ run from 0 to 3 and Latin indices $a$ and $b$ take values 1, 2, 3.
The particle distribution $\tilde{f}$ defined on the mass shell 
in this metric is given by 
$$  
\tilde{f}(t,x,p)=e^{-4\phi}f(t,x,e^\phi p). 
$$
More details on the derivation of this system are given in the next section. 
It should be emphasized that, although this model has no direct physical applications, scalar fields play a major role in modern
theories of classical and quantum gravity. For example, the Brans-Dicke gravitational theory \cite{BD}, which is continuously tested against
general relativity, is a combination of Einstein's and Nordstr\"om's theory.
The Nordstr\"om-Vlasov system is also interesting in a pure mathematical sense. A hope is that by studying this model one may
reach a better understanding of a class of systems consisting of hyperbolic and transport equations. 

However in order to justify this model as a {\it genuine} relativistic generalization of the (gravitational) Vlasov-Poisson system, it
is necessary to indicate the relation between the solutions of the two systems. 
The main goal of this paper is to prove that in the {\it non-relativistic} limit $c\to\infty$
the solutions of (\ref{NV1})-(\ref{NV2}) converge
to solutions of (\ref{vlasovpoisson1})-(\ref{vlasovpoisson2}) in a pointwise sense.  
The analogous result was proved in \cite{Sch1} for the relativistic Vlasov-Maxwell system (see \cite{LE} for the case of two space 
dimensions)
and in \cite{RR2} for the Einstein-Vlasov system with spherical symmetry (in the latter case a weaker form of convergence holds also in
the absence of symmetries, see \cite{Re}).  

This paper proceeds as follows. In section \ref{derivation} we provide a formal derivation of the Nordstr\"om-Vlasov system and state
our main results in full detail. The first of such results is a local existence theorem of solutions of the Nordstr\"om-Vlasov system
in an interval of time independent of the speed of light, which is a necessary step to proceed further in the study of the 
non-relativistic limit. The solution of the latter problem is our second result. 
Our analysis follows \cite{Sch1} to a large extent and is based on the use of certain representation formulae for the solutions of the
Nordstr\"om-Vlasov system which have been introduced in \cite{CaRe} and which will be adapted to the present case in section
\ref{preliminaries}. There we shall also prove some estimates needed in the sequel. One of these estimates states that the
distribution function $f$ is uniformly bounded, which permits to improve the conditional global existence result, Theorem 1
in \cite{CaRe}. (The general $L^q$ estimates on $f$ are proved in \cite{CaRe1}, where they are used to establish existence
of global weak solutions to the Nordstr\"om-Vlasov system). In section 4 we prove our main results.

\section{Derivation of the Nordstr\"om-Vlasov system and main results} \label{derivation}

We shall refer to the Nordstr\"om-Vlasov system as the set of equations which  
models the kinetic motion of a self-gravitating ensemble of collisionless particles  
in accordance to a gravitational theory satisfying the following assumptions: 
\begin{enumerate}[(1)] 
\item The gravitational forces are mediated by a scalar field $\phi$ and  
	  the effect of such forces is to conformally rescale the metric of the  
	  (four dimensional) spacetime according to the relation 
	  \begin{equation}\label{metric}
	  g=A^2(\phi)\eta, 
	  \end{equation} 
	  where $\eta$ is the Minkowski metric and $A$ is a positive function. 
\item Scale invariance property: There exists a one-parameter symmetry group 
	  whose action consists in rescaling $A(\phi)$ by a constant factor. 
\item Postulate of simplicity: The dynamics of the field $\phi$ is governed 
	  by second order differential equations. 
\item The matter (by which we mean any non-gravitational field) is universally  
	  coupled to the metric (\ref{metric}). 
\end{enumerate} 
It was observed in \cite{C} (appealing to the more general case of Scalar-Tensor  
theories considered in \cite{DE}) that the above assumptions single out 
a unique one-parameter family of scalar gravitation theories. This parameter  
appears because of the scale invariance property, which forces the conformal factor  
to be of the form $A(\phi)=\exp(\kappa\phi)$, with $\kappa>0$. Hence in this theory  
the spacetime is a Lorentzian manifold endowed with the metric 
\begin{equation} \label{metric.sc.inv} 
g=e^{2\kappa\phi}\eta. 
\end{equation} 
To write down the field equation of this scalar gravitation theory in a simple form,  
let us consider a system of Cartesian coordinates $\{x^0:=ct,\,x^1,\,x^2,\,x^3\}=\{x^\mu\}$,  
$c$ denoting the vacuum speed of light in Galilean frames. 
In these coordinates, the equation for $\phi$ takes the form 
\begin{equation} \label{field.eq} 
-c^{-2}\partial^2_t\phi + \Delta_x\phi = -4\pi \frac{G_*}{c^4}\kappa\, e^{4\kappa\phi}T. 
\end{equation} 
Here $G_*$ is a dimensional constant (the bare gravitational constant) and $T$ is  
the trace of the stress-energy tensor of the matter with respect to the physical metric $g$. 
In \cite{C} $c$, $G_*$ and $\kappa$ have been set equal to unity and the factor $4\pi$
has been removed for simplicity. As we already mentioned in the introduction, 
this scalar gravitation theory corresponds to the one considered in
\cite{EF, No}.  
 
In the case of the Nordstr\"om-Vlasov system, the dynamics of the matter is 
described by a non-negative, real-valued function  
$\tilde{f}$ which gives the probability density to find a particle in a given  
spacetime position $x^{\mu}$ and with a given four momentum $p^{\mu}$. 
We assume for simplicity that there is only one species of particle and choose units  
such that the proper mass of each particle is equal to one.  
The particle distribution $\tilde{f}$ is defined on the mass shell of  
the metric (\ref{metric.sc.inv}), which is the subset of the tangent bundle of  
spacetime defined by the condition $g_{\mu\nu}p^{\mu}p^{\nu} = -c^2, \, p^0 > 0$.  
This implies 
\begin{equation}\label{pnot} 
p^0 = \sqrt{e^{-2\kappa\phi}c^2+\delta_{ab} p^a p^b}. 
\end{equation}  
Using $(x^\mu,p^a)$ as coordinates on the mass shell and denoting by $dp$ the volume element $dp^1 dp^2 dp^3$,  
the stress-energy tensor for this matter model is 
$$ 
T^{\mu\nu} = -c\int \sqrt{|\det g|} \frac{p^{\mu}p^{\nu}}{p_0} \tilde{f}\,dp, 
$$  
which implies  
\begin{equation}\label{trace} 
T=-c^3\,e^{2\kappa\phi}\int\frac{\tilde{f}}{p^0}\,dp. 
\end{equation}   
Finally, the coupling between the scalar gravitational field and the matter is  
completed by requiring that the distribution $\tilde{f}$ of particles on  
the mass-shell is constant on the geodesics of the metric (\ref{metric.sc.inv}).  
This leads to the Vlasov equation: 
\begin{equation}\label{vlasov.eq} 
c^{-1}\partial_{t}\tilde{f} + \frac{p^a}{p^0} \partial_{x^a}\tilde{f} 
	     - \frac{p^\mu p^\nu}{p^0} \Gamma_{\mu\nu}^a \partial_{p^a}\tilde{f} = 0, 
\end{equation} 
where  
$\Gamma_{\mu\nu}^{\sigma} 
  = \kappa(\delta_{\nu}^{\sigma}\partial_{\mu}\phi 
          +\delta_{\mu}^{\sigma}\partial_{\nu}\phi 
		  -\eta_{\mu\nu}\partial^{\sigma}\phi)$  
are the Christoffel symbols of the metric (\ref{metric.sc.inv}). 
 
Our goal is to relate the solutions of the system (\ref{field.eq})--(\ref{vlasov.eq})  
to the solutions of the Vlasov-Poisson system (\ref{vlasovpoisson1})-(\ref{vlasovpoisson2})
satisfying the condition $\lim_{|x|\to\infty}U=0$ (isolated solutions). 
Hence $\big(f_\infty,U\big)$ solves the system 
\begin{align} 
&\partial_t f_\infty+p\cdot\nabla_xf_\infty-\nabla_xU\cdot\nabla_pf_\infty=0, \label{VP1}\\ 
&U= -G\int \frac{\rho_\infty (t,y)}{|y-x|}\, dy,			   							\label{VP2}\\ 
&\rho_\infty (t,x)=\int_{\mathbb{R}^3} f_\infty(t,x,p)\, dp,				\label{VP3} 
\end{align}  
where $G$ denotes the Newtonian gravitational constant which had been set equal to unity in (\ref{vlasovpoisson1})-(\ref{vlasovpoisson2}).
In order to get some light on the relation between the two systems, let us consider a formal expansion of the 
solutions of the Nordstr\"om-Vlasov system in powers of $1/c$: 
\begin{align*} 
\phi &= \phi_{0}+ c^{-1}\phi_1+ c^{-2}\phi_2+...\\ 
f &= f_0+ c^{-1}f_1+ c^{-2}f_2+... 
\end{align*}  
Substituting these into (\ref{field.eq}) and comparing the terms of the same order we obtain 
\begin{gather} 
\Delta_x\phi_{0}=0,\quad\Delta_x\phi_1=0, \label{expand1}\\ 
-\partial_t^2\phi_{0}+\Delta_x\phi_2=4\pi G_*\kappa e^{7\kappa\phi_0}\int f_0\, dp. \label{expand2} 
\end{gather} 
Assuming fields vanishing at infinity, (\ref{expand1}) implies $\phi_{0}=\phi_1=0$ 
and so (\ref{expand2}) reduces to (\ref{VP2}) 
with the identification $\phi_2\sim U,\,f_0\sim f_\infty$, provided that $G_* \kappa =G$. 
The latter condition, which is 
necessary in order to obtain the correct Newtonian limit, shows that the role of the scale invariance parameter $\kappa$ is merely the one 
of fixing the units of the corresponding theory. {\it We shall henceforth set $\kappa=G_*=G=1$ for simplicity}.  
 
To put the above formal discussion in a more rigorous mathematical context, we first rewrite the  
equations  
(\ref{field.eq})--(\ref{vlasov.eq}) with the ``unphysical'' particle density  
as in the formulation of \cite{CaRe}, namely  
$$f(t, x, p) = e^{4\phi} \tilde{f}(t, x, e^{-\phi} p).$$  
In this frame, the unknown $(f, \phi)$ satisfies the equations (\ref{NV1})-(\ref{NV2}).
We supply this system with initial data $0\leq f(0,x,p)=f^{\rm in}(x,p)$,
$\phi(0,x)=\phi_{0}^{\rm in}(x)$, $\partial_t\phi(0,x)=\phi_1^{\rm in}(x)$.

The following notation will be used. Given two functions $g$ and $h$ on $\mathbb{R}^n$ we write $g\lesssim h$ if the 
estimate $g\leq D h$ holds for a non-negative constant $D$ independent of $c\geq 1$.
The constant $D$ may also depend on the length of some time interval $[0, T]$, in which case we write
$g \lesssim h$ for $t \in [0, T]$. 
Furthermore we write 
$$ 
A=B+\mathcal{O}(c^{-\delta}), \quad \delta\geq 1, 
$$ 
if $|A(y)-B(y)|\lesssim \,c^{-\delta},\,\forall y\in\mathbb{R}^n$. 
We also set
\begin{equation}
\mathcal{P}_c(t) = \sup_{0\leq s<t} \{|p| : (x,p) \in \supp f(s) \}+1,
\end{equation}
where $\supp f(t)$ means the support of $f(t, x, p)$ on $(x, p) \in \mathbb{R}^6$ for each $t$.

Here are the main results of this paper:

\vspace{0.2cm}
\begin{theorem}\label{globalclassical}
Initial data $f^{\rm in}\in C_c^1(\mathbb{R}^6),\,\phi_{0}^{\rm in}\in C^{3}_b(\mathbb{R}^3),
\,\phi_1^{\rm in}\in C^2_b(\mathbb{R}^3)$
launch a unique classical solution $(f,\phi)\in C^1([0,T_\mathrm{max})\times \mathbb{R}^6)\times C^2([0,T_\mathrm{max})\times
\mathbb{R}^3)$ to the Cauchy problem for the Nordstr\"om-Vlasov
system (\ref{NV1})-(\ref{NV2}) in a maximal interval of time $[0,T_\mathrm{max})$. 
If $\mathcal{P}_c(T_\mathrm{max})<\infty$, 
then $T_\mathrm{max}=\infty$, i.e., the solution is global. 
\end{theorem}

\vspace{0.2cm}
Note that under the assumptions of Theorem \ref{globalclassical} the local time of existence 
may shrink to zero as the speed of light tends to $+\infty$. To remove
this possibility we specify more restrictive initial data:

\vspace{0.2cm}
\begin{theorem}\label{localexistence}
Assume $f^{\rm in}\in C_c^3(\mathbb{R}^6)$ and $\phi_{0}^{\rm in}=c^{-2}\g,\,\phi_1^{\rm in}=c^{-2}\h$, where 
\begin{equation*} 
\g (x)=- \iint \frac{f^{\rm in}(y,p)}{|y-x|}\,dp\,dy,\quad \h \in C^2_c(\mathbb{R}^3).
\end{equation*}
Corresponding to these data there exists a unique solution $(f,\phi)\in C^1([0,T)\times\mathbb{R}^6) \times C^2([0,T)\times\mathbb{R}^3)$ 
of 
(\ref{NV1})-(\ref{NV2}) in an interval of time $[0,T)$ independent of $c$ such that $\mathcal{P}_c(T)\lesssim 1$.
\end{theorem}

\vspace{0.2cm}
Finally we give the conditions under which solutions to the Nordstr\"om-Vlasov system converge in a pointwise sense to solutions of the
Vlasov-Poisson system in the non-relativistic limit.
\vspace{0.2cm}
\begin{theorem}\label{newlim}
Let the data for the Nordstr\"om-Vlasov system be given as in Theorem \ref{localexistence} and assume that 
\begin{itemize}
\item[($\star$)]
There exists a unique solution $(f,\phi)\in C^1 \big([0,T)\times\mathbb{R}^6 \big)
\times C^2 \big([0,T)\times\mathbb{R}^3 \big)$ of 
(\ref{NV1})-(\ref{NV2}) in an interval $[0,T)\subseteq [0,+\infty)$ independent of $c$ and the estimate 
$\mathcal{P}_c(T')\lesssim 1$ holds for all $0\leq T'<T$.
\end{itemize} 
Denote by $f_\infty\in C^1([0,\infty)\times\mathbb{R}^6)$ the global solution of (\ref{VP1})--(\ref{VP3}) 
with data $f^{\rm in}$, which is known to exist by \cite{Pf}.  
Then for every $T' \in [0, T)$ and $t\in [0,T']$:
\begin{gather} 
\partial_t\phi(t)=\mathcal{O}(c^{-1}),\quad c^{2}\phi(t)=U(t)+\mathcal{O}(c^{-1}),\quad
c^{2}\nabla_x\phi(t)=\nabla_xU(t)+\mathcal{O}(c^{-1}),\label{newlimit1}\\ 
f(t)=f_\infty(t)+\mathcal{O}(c^{-1}). \label{newlimit2}
\end{gather} 
\end{theorem} 

In the notation of the spaces of functions used above, the subscript $c$ 
indicates that functions are compactly supported and $b$ means that all
the derivatives up to the indicated order are bounded. 
Theorem \ref{globalclassical} will be proved as a corollary of Theorem 1 of
\cite{CaRe} in the next section and Theorems \ref{localexistence} and \ref{newlim} will be proved in Section \ref{proof}.

To conclude this section, we remark that more general data for the field are allowed in Theorems \ref{localexistence} and \ref{newlim}. For instance we may
require that $\phi_0^{\mathrm{in}}=c^{-2} \g + c^{-3}g^\flat$, where $g^\flat\in C^3_b(\mathbb{R}^3)$. This modifies our estimates only by
terms which are of higher order in powers of $c^{-1}$,  without affecting the general argument. In order to make the estimates below more
transparent, we take for simplicity $g^\flat\equiv 0$.  

\section{Preliminaries and proof of Theorem \ref{globalclassical}} \label{preliminaries}

First note that the classical solution of (\ref{NV1}) is  
\begin{align}
\phi(t,x) &= \phi_{\rm hom}(t,x) 
		     - \frac{1}{c^2} \intL\int\frac{f\farg}{\sqcp |y-x|}\,dp\,dy \notag\\
   		  &:=\phi_{\rm hom}(t, x)+\psi(t, x),    \label{phi} 
\end{align}  
where
\begin{equation}\label{phi0} 
\phi_{\rm hom}(t,x) = \partial_t\Big(\frac{t}{4\pi}\intw \phi_{0}^{\rm in}(x+ct\omega)\,d\omega\Big) 
			+ \frac{t}{4\pi} \intw \phi_1^{\rm in}(x+ct\omega) \,d\omega 
\end{equation} 
is the solution of the homogeneous wave equation with data $\phi_0^{\rm in}$ and $\phi_1^{\rm in}$ 
and $\psi$ the solution of (\ref{NV1}) with trivial data. We start with estimating the homogeneous part of the field $\phi$.
\begin{lemma}\label{esthom}
Let the initial data for the field be given as in Theorems \ref{localexistence} and \ref{newlim}. Then we have
\begin{equation*}
\|\phi_{\rm hom} (t) \|_\infty \lesssim c^{-1} (1+t).
\end{equation*}
\end{lemma}

\noindent{\it Proof:} By means of (\ref{phi0}) and the assumptions on the data $\phi_0^{\rm in}$ and $\phi_1^{\rm in}$  
we have
\begin{equation}\label{phihom}
\phi_{\rm hom}(c^{-1}t,x) = \frac{1}{c^2}\left[\partial_t\Big(\frac{t}{4\pi}\intw \g(x+t\omega)\,d\omega\Big) 
			+ \frac{c^{-1}t}{4\pi} \intw \h(x+t\omega) \,d\omega\right]. 
\end{equation}
The term in the square brackets in (\ref{phihom}) is estimated by $D(1+t)$
where 
\[
D:=\|g^\sharp\|_\infty+\|\nabla_x g^\sharp\|_\infty+\|h^\sharp\|_\infty.
\] 
Hence $\|\phi_{\rm hom}(c^{-1}t)\|_\infty \lesssim c^{-2}(1+t)
$, 
which implies $\|\phi_{\rm hom} (t) \|_\infty \lesssim c^{-1} (1+t).$
 \prfe

The following estimate is crucial for extending the argument of \cite{Sch1} to the Nordstr\"om-Vlasov system.
\begin{proposition}\label{boundf}
The distribution function $f$ satisfies the estimate
$$
\|f(t)\|_{\infty}\leq \|f^{\rm in}\|_\infty \exp\Big[4\big(\|\phi_{\rm hom}(t)\|_{\infty}+\|\phi^{\rm in}_0\|_{\infty}\big)\Big].
$$
In particular, for data as in Theorems \ref{localexistence} and \ref{newlim}, we have $\|f(t)\|_{\infty}\lesssim e^{4Dt}$ for all $t\in\mathbb{R}$.
\end{proposition}

\noindent{\it Proof :} Let $(X,P)(s,t,x,p)$ denote the characteristics of (\ref{NV2}) which satisfy the condition $(X,P)(t,t,x,p) = (x,p)$.
In short, we use $X(s) := X(s, t, x, p)$ and $P(s) := P(s, t, x, p)$ instead.
Note that the function $e^{-4\phi}f$ is constant along these curves. Hence the solution of (\ref{NV2}) is given by 
\begin{align}
f(t,x,p)&= f^{\rm in}(X(0),P(0))
\exp \left[4\phi(t,x)-4\phi^{\rm in}(X(0))\right]\nonumber\\
&= f^{\rm in}(X(0),P(0))\exp \left[-4\phi^{\rm in}(X(0))\right]\exp \left[4\phi_{\rm hom}(t,x)\right]
\exp\left[4\psi(t,x)\right].\label{frep}
\end{align}
Since $\psi\leq 0$, then $e^{4\psi}\leq 1$ and the claim follows.\prfe

Combining this result with the one in \cite{CaRe} we obtain the following.

\noindent{\it Proof of Theorem \ref{globalclassical}:} 
It is enough to prove the theorem for $c=1$. For given a solution $(f,\phi)$ of 
(\ref{NV1})-(\ref{NV2}), then the solution obtained by the rescaling $c f(c^{-1}t,x,cp)$, $\phi(c^{-1}t,x)$
solves the system with $c=1$. The claim has been proved in \cite{CaRe} under the additional condition that
$\mathcal{Q}(T_\mathrm{max})<\infty$, where
$$
\mathcal{Q}(t)=\sup_{0\leq s<t}\{|\phi(t,x)|:(x,p)\in\textnormal{supp}f(s)\}.
$$
Now, assuming compact support in $p$ for the distribution function, i.e., 
$\mathcal{P}_1(T_\mathrm{max})<\infty$, 
it follows by Proposition \ref{boundf} that the right hand side of (\ref{phi}) is bounded in $L^\infty$.
Hence $\phi$ itself is bounded and thus the condition $\mathcal{Q}(T_\mathrm{max})<\infty$ is satisfied. 
The claim follows by Theorem 1 of \cite{CaRe}. \prfe 

Next we derive the representation formulae for the first order derivatives of the field as in 
\cite{CaRe}, but for arbitrary values of $c$ and also with all data terms specified. 

One can see by the Vlasov equation (\ref{NV2}) that 
$$ 
Sf = \{(\partial_t \phi)p +c^2(\cp)^{-1/2} \nabla_x \phi\} \cdot \nabla_p f 
	+ 4  (S\phi)f. 
$$  
From (\ref{phi}) we have 
\begin{align*} 
\partial_t \phi(t, x) &= \partial_t \phi_{\rm hom}(t, x) -  c^{-2}t^{-1} 
		   		   \intE \int \frac{f^{\rm in}(y, p)}{\sqcp} \,dp \,dS_y\\ 
				&\quad   -  c^{-2} \intL \int  
					  \frac{\partial_t f\farg}{\sqcp |y-x|} \,dp \,dy. 
\end{align*} 
Now using the identity 
\begin{align*} 
\partial_tg\farg &= (\op)^{-1}\big\{(Sg)\farg\notag \\ 
  				 &\quad-\widehat{p} \cdot \nabla_y[g\farg] \big\} 
\end{align*} 
and integration by parts we achieve the following representation for $\partial_t\phi$. 
 
\begin{proposition}\label{Prop_phi_t} 
\begin{align*} 
&\partial_t \phi(t,x) 
 =  \partial_t \phi_{\rm hom}(t, x) \\ 
&\qquad -   c^{-2}t^{-1} \intE \int \frac{f^{\rm in}(y,p)}{(\op)\sqcp} \,dp \,dS_y\\ 
&\qquad -  c^{-2} \intL \int a^{\phi_t}(\omega, p) f\farg \,dp \frac{dy}{|y-x|^2}\\ 
&\qquad -  c^{-2} \intL \int b^{\phi_t}(\omega, p)  
	   	 	  	  		   (S\phi) f\farg \,dp\frac{dy}{|y-x|}\\ 
&\qquad - c^{-1} \intL \int c^{\phi_t}(\omega, p)  
		  		   		  (\nabla_x\phi) f\farg \,dp\frac{dy}{|y-x|}, 
\end{align*} 
where the kernels are 
\begin{align*} 
a^{\phi_t}(\omega, p) &= - \frac{\widehat{p}\cdot(\oap)}{(\op)^2\sqcp},\\ 
b^{\phi_t}(\omega, p) &= \frac{(\oap)^2}{(\op)^2\sqcp},\\ 
c^{\phi_t}(\omega, p) &= \frac{\oap}{(\op)^2(\cp)^{3/2}}							   
\end{align*} 
and $\omega=(y-x)/|y-x|$. 
\end{proposition}  
  
The process to obtain the representation for $\nabla_x\phi$ is similar to  
the way for $\partial_t\phi$, but now with the following identity:  
\begin{align*} 
\partial_{y_i}g\farg &= c^{-1}\omega_i (\op)^{-1}(Sg)\farg \\ 
 				 &\, + \Big(\delta_{ik}-\frac{c^{-1}\omega_i\widehat{p}_k}{\op}\Big)  
     			   		  \partial_{y_k}[g\farg],\, i=1,2,3. 
\end{align*}
\begin{proposition}\label{Prop_phi_x} 
The representation of $\partial_{x_i}\phi$ is 
\begin{align*} 
& \partial_{x_i} \phi(t,x) =  \partial_{x_i} \phi_{\rm hom}(t, x)\\ 
&\qquad -  c^{-3}t^{-1} \intE \int \frac{\omega_i}{(\op)\sqcp} f^{\rm in}(y,p) \,dp \,dS_y\\ 
&\qquad - c^{-3} \intL \int a^{\phi_{x_i}}(\omega, p) f\farg \,dp \frac{dy}{|y-x|^2}\\ 
&\qquad - c^{-3} \intL \int b^{\phi_{x_i}}(\omega, p)  
	   	 	  	  		   (S\phi) f\farg \,dp \frac{dy}{|y-x|}\\ 
&\qquad -  c^{-2} \intL \int c^{\phi_{x_i}}(\omega, p)  
	   	 	  	  		   (\nabla_x\phi) f\farg \,dp \frac{dy}{|y-x|}, 
\end{align*} 
where the kernels are 
\begin{align*} 
a^{\phi_{x_i}}(\omega, p) &= \frac{c(\oap)_i - c^{-1}(\widehat{p}\wedge(\owp))_i}{(\op)^2 \sqcp},\\ 
b^{\phi_{x_i}}(\omega, p) &= \omega_i b^{\phi_t},\\ 
c^{\phi_{x_i}}(\omega, p) &= \omega_i c^{\phi_t}.							   
\end{align*} 
\end{proposition}

\section{Proof of Theorems \ref{localexistence} and \ref{newlim}}\label{proof} 

In this section we prove our main results. We shall frequently use Lemmas 1 and 2 of \cite{Sch1}, 
which we state below for future reference.
\begin{lemma}\label{lemma1}
For all $g \in C^0_c(\mathbb{R}^3)$, we have 
$$
\xi\intw |g(x+\xi\omega)|\,d\omega\lesssim 1,
$$
for $\xi \geq 0$.
\end{lemma}
\begin{lemma}\label{lemma2}
Let $h\in C^{2}(\mathbb{R}^3)$ such that $\Delta h\in C^0_c(\mathbb{R}^3)$. Then for $c>0$ and $t\geq 0$,
$$
\partial_t\Big( t \intw h(x+ct\omega)\,d\omega\Big)=-\intG\frac{\Delta h(y)}{|y-x|}\,dy.
$$ 
\end{lemma}

The next Lemma contains two simple estimates which are often used in the sequel.
\begin{lemma}\label{lemma3}
\begin{itemize}
\item[(i)] $(\op)^{-1}\lesssim \mathcal{P}_c(t)^2$, for $(x,p)\in \supp f(t)$
\item[(ii)] $f(t,x,p)=0$, for $|x|\geq R+\mathcal{P}_c(t)t$, where $R := \sup \{|x| : (x, p) \in \supp f^{\rm in} \} $.
\end{itemize}
\end{lemma}  

\noindent\textit{Proof:} By $|p| \leq \mathcal{P}_c(t)$,
$$
\op \geq 1-\frac{|p|}{\sqrt{c^2+p^2}} = \frac{c^2}{\sqrt{c^2+p^2}\big(\sqrt{c^2+p^2} + |p|\big)}
\geq \frac{c^2}{2 \big(c^2+\mathcal{P}_c(t)^2\big)},
$$
by which (i) follows. The property (ii) on the support of $f$ follows by
(\ref{frep}) and the definition of characteristics.\prfe
\noindent{\it Proof of Theorem \ref{localexistence}}: 
From Proposition \ref{Prop_phi_x}, we have 
\begin{align}
&\partial_{x_i} \phi(t,x) =  \partial_{x_i} \phi_{\rm hom}(t, x)\notag \\
&\qquad -  c^{-3}t^{-1} \intE \int \frac{\omega_i}{(\op)\sqcp} f^{\rm in}(y,p) \,dp \,dS_y \notag\\
&\qquad + I_{x_i} + II_{x_i} + III_{x_i},\label{phi_xi}
\end{align}
where
\begin{equation}\label{c2dx_phihom}
\partial_{x_i} \phi_{\rm hom}(t,x) 
= c^{-2}\partial_t\Big(\frac{t}{4\pi}\intw \partial_{x_i}\g (x+ct\omega)\,d\omega\Big) 
			+ c^{-2}\frac{t}{4\pi} \intw \partial_{x_i} \h (x+ct\omega) \,d\omega.
\end{equation}
By the assumption $\h  \in C^2_c(\mathbb{R}^3)$, Lemma \ref{lemma1} gives
\begin{equation*}
\left|\frac{c^{-2}t}{4 \pi} \intw \partial_{x_i} \h (x + ct\omega)\,d\omega\right| \lesssim c^{-3}.
\end{equation*}
For the first term in (\ref{c2dx_phihom}), using Lemma \ref{lemma2}
and the identity $\Delta \g  = 4 \pi \int f^{\rm in} \, dp$ we get
\begin{align*}
 \partial_t \Big( \frac{t}{4 \pi} \intw \partial_{x_i}\g (x+ct\omega)\,d\omega \Big)
&= - \partial_{x_i}\intG \int \frac{f^{\rm in}(y, p)}{|y-x|} \,dp \,dy  \notag\\
& = (ct)^{-1} \intE \int \omega_{i}f^{\rm in}(y, p) \, dp \,dS_y \notag\\
&\qquad - \intG  \int \frac{\omega_i f^{\rm in}(y, p)}{|y-x|^2} \,dp \,dy.
\end{align*}
The surface integral will be combined with the second term in (\ref{phi_xi}). First note that, using (i) in Lemma \ref{lemma3} and
$|\widehat{p}|\leq\mathcal{P}_c(t)$,  
\begin{align}
&\left| 1 - \frac{1}{(\op) \sqcp} \right|\notag\\
& \quad \leq \left|1 - \frac{1}{\sqcp} \right| + \left| \frac{1}{\sqcp} - \frac{1}{(\op) \sqcp}  \right|\notag\\
& \quad \lesssim c^{-2}\mathcal{P}_c(t)^2 + \frac{1}{\sqcp}\left[\frac{c^{-1}|\wdp|}{(\op)}\right]
\lesssim c^{-1}\mathcal{P}_c(t)^3. \label{kernel1} 
\end{align} 
So we get
\begin{align*}
&\left|c^{-3}t^{-1} \intE \int \omega_i f^{\rm in}(y, p) \, dp \, dS_y \right.\notag\\
&\qquad\qquad \left.- c^{-3}t^{-1} \intE \int \frac{\omega_i f^{\rm in}(y, p) \, dp \, dS_y}{(\op) \sqcp} \right| \notag\\
&\qquad\qquad\qquad \lesssim c^{-3} t^{-1} \intE \int c^{-1}\mathcal{P}_c(t)^3|\omega_i| f^{\rm in}(y, p) \, dp \, dS_y \notag\\
&\qquad\qquad\qquad \lesssim c^{-3} \mathcal{P}_c(t)^3\Big (ct \intw \int f^{\rm in}(x+ct\omega, p) \, dp \, d\omega \Big)
\lesssim c^{-3}\mathcal{P}_c(t)^3. 
\end{align*}
Split the kernel in $I_{x_i}$ according to 
\begin{equation*}
\tilde{a}^{\phi_{x_i}} (\omega, p) 
:= \frac{\widehat{p}_i - c^{-1}\big( \widehat{p} \wedge (\owp)\big)_i}{(\op)^2 \sqcp} 
= a^{\phi_{x_i}}(\omega,p)-\frac{c\,\omega_i}{(\op)^2 \sqcp}.
\end{equation*}
By (i) in Lemma \ref{lemma3} one can see that
$|\tilde{a}^{\phi_{x_i}}| \lesssim \mathcal{P}_c(t)^5$. Using Proposition \ref{boundf} and (ii) of Lemma \ref{lemma3} we obtain
\begin{align*}
&\left| - c^{-3} \intL \int \tilde{a}^{\phi_{x_i}}(\omega, p) f\farg \, dp \frac{dy}{|y-x|^2}\right|\\ 
& \qquad \leq c^{-3} \int_{|y| \leq R + \mathcal{P}_c(t)t} \int_{|p| \leq \mathcal{P}_c(t)}
  		 	  		 |\tilde{a}^{\phi_{x_i}}(\omega, p)| f\farg \, dp \frac{dy}{|y-x|^2} \notag\\
& \qquad \lesssim c^{-3} \sup_{0\leq\tau\leq t}\|f(\tau)\|_\infty \mathcal{P}_c(t)^5
					 \int_{|y| \leq R + t\mathcal{P}_c(t)} \int_{|p| \leq \mathcal{P}_c(t)}
  		 	  		 	 	  			  \, dp \frac{dy}{|y-x|^2} \notag\\
&\qquad\lesssim c^{-3}\mathcal{P}_c(t)^9 (1+t)\,e^{4Dt}.\notag
\end{align*}
A computation similar to (\ref{kernel1}) shows that
\begin{equation*}
\left| 1 - \frac{1}{(\op)^2 \sqcp} \right|
\lesssim c^{-1}\mathcal{P}_c(t)^5. 
\end{equation*}
Hence 
\begin{align*}
&\left| c^{-2} \intL \int \omega_i f\farg \, dp \frac{dy}{|y-x|^2} \right. \notag\\
&\quad \left. - c^{-2} \intL \int \frac{\omega_i}{(\op)^2\sqcp}  f\farg \, dp \frac{dy}{|y-x|^2} \right| \notag\\
& \qquad \lesssim c^{-2} \int_{|y| \leq R + \mathcal{P}_c(t)t} \int_{|p| \leq \mathcal{P}_c(t)}  
c^{-1}\mathcal{P}_c(t)^5 |\omega_i|f\farg \, dp \frac{dy}{|y-x|^2} \notag\\
&\qquad\lesssim c^{-3}\mathcal{P}_c(t)^9 (1+t)\, e^{4Dt}. 
\end{align*}
Combining the estimates obtained thus far we get
\begin{align}
&\left|\partial_{x_i}\phi_{\rm hom}(t,x) -  c^{-3}t^{-1} \intE \int \frac{\omega_i}{(\op)\sqcp} f^{\rm in}(y,p) \,dp \,dS_y
+I_{x_i}\right|\notag\\ 
&\lesssim\left| -c^{-2} \iint \omega_i f(\max\{0, t-c^{-1}|y-x|\}, y, p) \, dp \frac{dy}{|y-x|^2}\right| + c^{-3}\mathcal{P}_c(t)^9 e^{4Dt}\notag\\
&\lesssim c^{-2}\mathcal{P}_c(t)^3 \sup_{[0, t]}\|f(s)\|_\infty \int_{|y|\leq
R+\mathcal{P}_c(t)t}\frac{dy}{|y|^2}+c^{-3}\mathcal{P}_c(t)^9 e^{4Dt}\notag\\ 
&\lesssim c^{-2}\mathcal{P}_c(t)^9(1+t)e^{4Dt}\label{leadingorder}.
\end{align}
Now we estimate  $II_{x_i}$  and $III_{x_i}$. 
Again by (i) of Lemma \ref{lemma3} we obtain the bounds
$$ 
|b^{\phi_{x_i}}(\omega, p)|\lesssim \mathcal{P}_c(t)^4, 
\quad |c^{\phi_{x_i}}(\omega, p)|\lesssim \mathcal{P}_c(t)^4. 
$$ 
Let us define  
$$ 
K_c(t)=\sup \{c|\partial_t\phi(t,x)|+c^2|\nabla_x\phi(t,x)|,\, x\in \mathbb{R}^3\}. 
$$ 
Hence, using $c|S(\phi)|\leq K_c(t)$, we get  

\begin{align*} 
|II_{x_i}| 
&\lesssim 
 c^{-3} \mathcal{P}_c(t)^4 \int_0^t \int_{|y-x|=c(t-\tau)}\int_{|p|\leq \mathcal{P}_c(\tau)}f(\tau,y,p)|y-x|^{-1}K_c(\tau)\,dp\,dS_y\,d\tau\\ 
&\lesssim c^{-2}\mathcal{P}_c(t)^7 t e^{4Dt} \int_0^t K_c(\tau)d\tau. 
\end{align*}
$III_{x_i}$ satisfies an identical estimate, since $c^2|\nabla_x\phi|\leq K_c(t)$. Collecting the various bounds we obtain 
\begin{equation}\label{boundphix} 
|\nabla_x\phi(t,x)|\lesssim c^{-2}\mathcal{P}_c(t)^9 (1+t) e^{4Dt} \Big(1+\int_0^t K_c(\tau)d\tau\Big). 
\end{equation} 

Now the estimate on the time derivative of the field. 
From Proposition \ref{Prop_phi_t} we have 
\begin{align} 
\partial_t \phi(t,x) & =  \partial_t \phi_{\rm hom}(t, x)  
 -   c^{-2}t^{-1} \intE \int \frac{f^{\rm in}(y,p)}{(\op)\sqcp} \,dp \,dS_y \notag\\ 
&\quad + I_t + II_t + III_t, \label{2nd_data} 
\end{align} 
where
$$
\partial_t\phi_{\rm hom}(t,x) = \frac{1}{c}\left[\partial_t\Big(\frac{c^{-1}t}{4\pi}\intw \h(x+ct\omega)\,d\omega\Big) 
			+ \frac{ct}{4\pi} \intw \Delta \g(x+ct\omega) \,d\omega\right]. 
$$
A direct application of Lemma \ref{lemma1} now gives $\|\partial_t \phi_{\rm hom}(t)\|_\infty \lesssim c^{-1}(1+t)$. 
For the second term of (\ref{2nd_data}) we have
\begin{align*} 
& c^{-2}t^{-1} \intE \int \frac{f^{\rm in}(y,p)}{(\op)\sqcp} \,dp \,dS_y\\ 
&\quad \lesssim  c^{-1} \mathcal{P}_c(0)^2 \left[ ct \intw  \int_{|p|\leq \mathcal{P}_c(0)} f^{\rm in}(x+ct\omega, p) dp\, d\omega\right] 
	   \lesssim c^{-1}.
\end{align*} 
In order to estimate the remaining terms in (\ref{2nd_data}) we use the bounds 
$$ 
|a^{\phi_{t}}(\omega, p)|\lesssim \mathcal{P}_c(t)^5,\quad |b^{\phi_{t}}(\omega, p)|\lesssim \mathcal{P}_c(t)^4, 
\quad |c^{\phi_{t}}(\omega, p)|\lesssim \mathcal{P}_c(t)^4. 
$$ 
The estimate for $I_t$ follows: 
\begin{align*} 
|I_t| 
&\lesssim
c^{-2}\mathcal{P}_c(t)^5\intL\int_{|p|\leq 
\mathcal{P}_c(t)}f\farg\,dp\,\frac{dy}{|y-x|^2}\\ 
&\lesssim 
c^{-2}\mathcal{P}_c(t)^8 \sup_{[0, t]}\|f(s)\|_\infty \int_{|y|\leq R+\mathcal{P}_c(t)t}\frac{dy}{|y|^2} 
\lesssim 
c^{-2}\mathcal{P}_c(t)^9 (1+t) e^{4Dt}. 
\end{align*} 
Also
\begin{align*} 
|II_t| 
&\lesssim 
 c^{-2} \mathcal{P}_c(t)^4 \int_0^t \int_{|y-x|=c(t-\tau)}\int_{|p|\leq \mathcal{P}_c(\tau)}f(\tau,y,p)|y-x|^{-1}K_c(\tau)\,dp\,dS_y\,d\tau\\ 
&\lesssim c^{-1}\mathcal{P}_c(t)^7 t e^{4Dt} \int_0^t K_c(\tau)d\tau. 
\end{align*} 
$III_t$ satisfies an identical estimate, since $c^2|\nabla_x\phi|\leq K_c(t)$. Collecting the various bounds we obtain 
\begin{equation}\label{boundphit} 
|\partial_t\phi(t,x)|\lesssim c^{-1}\mathcal{P}_c(t)^9 (1+t) e^{4Dt} \Big(1+\int_0^t K_c(\tau) \,d\tau\Big). 
\end{equation} 
Combining (\ref{boundphit}) and (\ref{boundphix}) entails 
\begin{equation*} 
K_c(t)\lesssim \mathcal{P}_c(t)^9 (1+t) e^{4Dt} \Big(1+\int_0^t K_c(\tau)d\tau\Big). 
\end{equation*} 
Hence by Gronwall's inequality, 
\begin{equation}\label{boundK2} 
K_c(t)\lesssim  \mathcal{P}_c(t)^9 (1+t)e^{4Dt} \exp\Big(\mathcal{P}_c(t)^9(1+t)^2 e^{4Dt} \Big). 
\end{equation} 
Note that the characteristics $(X, P)(s)$ of (\ref{NV2}) with $(X, P)(t)=(x, p)$ satisfies
$$
\frac{dP}{ds}= -(S\phi)(s,X)\,P -\frac{c^2\nabla_x\phi(s,X)}{\sqrt{1+P^2}}.
$$
So 
\begin{equation}\label{bound_p}
|p| \lesssim |P(0)| + \int_0^t \mathcal{K}_c(\tau)\mathcal{P}_c(\tau) d\tau 
		   \lesssim \mathcal{P}_c(0)-1 +\int_0^t \mathcal{K}_c(\tau)\mathcal{P}_c(\tau) d\tau.
\end{equation}
Therefore by (\ref{boundK2}) and the definition of $\mathcal{P}_c(t)$, (\ref{bound_p}) becomes
\begin{equation}\label{bound_pc}
\mathcal{P}_c(t)\lesssim 1+\int_0^t(1+\tau)\mathcal{P}_c(\tau)^{10} e^{4D\tau}\exp\Big(\mathcal{P}_c(\tau)^9(1+\tau)^2 e^{4D\tau} \Big) d\tau.
\end{equation}
By Gronwall's inequality, there exists an interval $[0,T)$ independent of $c$ where 
$\mathcal{P}_c(t)$ remains finite for all $c \geq 1$, i.e., $\mathcal{P}_c(t)\lesssim 1$. Using this estimate we can complete the proof of
Theorem \ref{localexistence}. Let $T_\mathrm{max}^c$ denote the maximal time of existence of a solution of (\ref{NV1})-(\ref{NV2}) and
assume $T_\mathrm{max}^c <T$ for some $c\geq 1$. Since $\mathcal{P}_c(t)$ is an increasing function of time, this implies 
$\mathcal{P}_c(T_\mathrm{max}^c)<\infty$ and so, by Theorem \ref{globalclassical}, $T_\mathrm{max}^c=\infty$, a contradiction. Hence, the
solution is defined on the interval $[0,T)$ for all $c\geq 1$. 

\vspace{0.1cm}
\noindent{\it Proof of Theorem \ref{newlim}}: 
For brevity we omit stating that the estimates below are valid for $t\in [0,T']$. The claim on $\partial_t\phi$ follows directly by
(\ref{boundphit}) and the assumption $\mathcal{P}_c(t)\lesssim 1$. Note also that, by (\ref{boundphix}),
$\nabla_x\phi=\mathcal{O}(c^{-2})$.  
We start with estimating $\phi$. By the assumption we made on $\phi_1^{\rm in}$ and Lemma \ref{lemma1},
the second term in (\ref{phi0}) becomes
\begin{equation}\label{phi_data2}
\frac{t}{4\pi} \intw \phi_1^{\rm in}(x+ct\omega) \,d\omega = \mathcal{O}(c^{-3}).
\end{equation}
For the first term in (\ref{phi0}), with the assumption on $\phi_0^{\rm in}$, Lemma \ref{lemma2}
and the fact that $\Delta \g  = 4 \pi \int f^{\rm in} \, dp$, we get 
\begin{align}\label{phi_data1}
\partial_t\Big(\frac{t}{4\pi}\intw \phi_0^{\rm in}(x+ct\omega)\,d\omega\Big)
& = - (4 \pi)^{-1} c^{-2} \intG \frac{\Delta \g (y)}{|y-x|} \,dy \notag\\
& = - c^{-2} \intG \int \frac{f^{\rm in}(y, p)}{|y-x|} \,dp \,dy.
\end{align}
For $\psi(t, x)$ in (\ref{phi}), first consider
\begin{align*}
&\left| c^{-2} \intL\int \frac{f\farg}{|y-x|} - \frac{f\farg}{\sqcp |y-x|}\,dp\,dy \right| \\
& \qquad \lesssim c^{-2} \int_{|y| \leq R+\mathcal{P}_c(t)t} \int_{|p| \leq \mathcal{P}_c(t)} c^{-2} f\farg\,dp \frac{dy}{|y-x|} 
\lesssim c^{-4}.
\end{align*}
Then $\psi$ becomes
\begin{equation} \label{psi}
\psi(t, x) = - c^{-2} \intL\int f\farg\,dp \frac{dy}{|y-x|} + \mathcal{O}(c^{-4}). 
\end{equation}
So collecting (\ref{phi_data2}), (\ref{phi_data1}) and (\ref{psi}), we get
\begin{align}
\phi(t, x) & = - c^{-2} \intL\int f\farg\,dp \frac{dy}{|y-x|}\notag\\ 
& \qquad - c^{-2} \intG \int \frac{f^{\rm in}(y, p)}{|y-x|} \,dp \,dy + \mathcal{O}(c^{-3})\notag\\
& = - c^{-2} \iint f(\max\{0, t-c^{-1}|y-x|\}, y, p) \,dp \frac{dy}{|y-x|} + \mathcal{O}(c^{-3}).\label{forphi}
\end{align}
So using (\ref{VP2}) and (\ref{forphi}), now we estimate
\begin{align} 
|c^2 \phi(t, x) - U(t, x)| 
&= \left| \mathcal{O}(c^{-1}) + \iint f_\infty(t, y, p) \, dp \frac{dy}{|y-x|}\right. \notag\\
& \qquad \left.- \iint f(\max\{0, t-c^{-1}|y-x|\}, y, p) \, dp \frac{dy}{|y-x|}\right|\notag\\
& \lesssim \iint \Big| f_\infty(\max\{0, t-c^{-1}|y-x|\}, y, p) - f_\infty(t, y, p)\Big| \, dp \frac{dy}{|y-x|} \notag\\
& \qquad +\iint \Big| f(\max\{0, t-c^{-1}|y-x|\}, y, p) \notag\\
& \qquad\qquad- f_\infty (\max\{0, t-c^{-1}|y-x|\}, y, p)\Big|\, dp \frac{dy}{|y-x|} + c^{-1}.\label{phi_U}
\end{align}
Define 
\begin{equation*}
D_F(t) := \sup \{ |f(\tau, x, p) - f_\infty(\tau, x, p)| : \tau \in [0, t],\, x \in \mathbb{R}^3 \text{ and } p \in \mathbb{R}^3 \}.
\end{equation*}
Also define $\mathcal{P}_\infty(t)$ as the following
\begin{equation*}
\mathcal{P}_\infty(t) = \sup_{0\leq s<t} \{|p| : (x,p) \in \supp f_\infty(s) \}+1.
\end{equation*}
Since ($f_\infty$, $U$) is a $C^1$ solution of (\ref{VP1})--(\ref{VP3}) 
and the initial data $f_\infty$ has compact support,
$\mathcal{P}_\infty$ is well defined for all $t \geq 0$.
Note that $\partial_t f_\infty$ is bounded on $\mathbb{R}^6 \times [0, T']$.
Also let $\mathcal{P}(t) := \mathcal{P}_c(t) + \mathcal{P}_\infty(t)$.
Then (\ref{phi_U}) becomes
\begin{align}
& |c^2 \phi(t, x) - U(t, x)| \notag\\
& \qquad \lesssim \int_{|y|\leq R+\mathcal{P}(t)t} \int_{|p|\leq \mathcal{P}(t)} \int^t_{\max\{0, t-c^{-1}|y-x|\}}
  		 							 |\partial_t f_\infty(s, y, p)| \,ds \, dp \frac{dy}{|y-x|}\notag\\
& \qquad\qquad + \int_{|y|\leq R+\mathcal{P}(t)t} \int_{|p|\leq \mathcal{P}(t)} D_F(\max\{0, t-c^{-1}|y-x|\})\, dp \frac{dy}{|y-x|}  + c^{-1}. \notag\\
& \qquad \lesssim D_F(t) + c^{-1}. \label{phi_U_final}
\end{align}

Before estimating $D_F(t)$, let us look at $\nabla_x \phi$. In the process of proving  Theorem \ref{localexistence} we have also shown
that
\begin{eqnarray}
&\partial_{x_i} \phi_{\rm hom}
-  c^{-3}t^{-1} \intE \int \frac{\omega_i}{(\op)\sqcp} f^{\rm in}(y,p) \,dp \,dS_y + I_{x_i} \nonumber\\
&\quad =-c^{-2} \iint \omega_i f(\max\{0, t-c^{-1}|y-x|\}, y, p) \, dp \frac{dy}{|y-x|^2} + \mathcal{O}(c^{-3}),\label{phixi_I}
\end{eqnarray}
cf. (\ref{leadingorder}).
For $II_{x_i}$ in (\ref{phi_xi}) we use $S(\phi) = \mathcal{O}(c^{-1})$ and $|b^{\phi_{x_i}}|\lesssim \mathcal{P}_c(t)^4$. 
Therefore
\begin{equation}\label{phixi_II}
|II_{x_i}| \lesssim c^{-3} \int_{|y| \leq R + \mathcal{P}_c(t)t} \int_{|p| \leq \mathcal{P}_c(t)} 
		   c^{-1} f\farg \,dp \frac{dy}{|y-x|}
		   \lesssim c^{-4}. 
\end{equation}

The estimation of $III_{x_i}$ is similar to the one of $II_{x_i}$. 
Recall that $|c^{\phi_{x_i}}|\lesssim \mathcal{P}_c(t)^4$ and $\nabla_x \phi = \mathcal{O}(c^{-2})$.
So we get
\begin{equation} \label{phixi_III}
|III_{x_i}| \lesssim c^{-4}.
\end{equation}

Now collecting (\ref{phixi_I})--(\ref{phixi_III}) we obtain
\begin{equation}
\partial_{x_i} \phi 
= - c^{-2} \iint (y_i - x_i) f(\max\{0, t-c^{-1}|y-x|\}, y, p) \, dp \frac{dy}{|y-x|^3} +\mathcal{O}(c^{-3}). \label{xi_phi}
\end{equation}
By the similar argument in (\ref{phi_U_final}), with (\ref{VP2}) and (\ref{xi_phi}), we estimate
\begin{align}
& |c^2 \nabla_x \phi(t, x) - \nabla_x U(t, x)| \notag\\
&\qquad = \left| \mathcal{O}(c^{-1}) + \iint (y-x) f_\infty(t, y, p) \, dp \frac{dy}{|y-x|^3} \right. \notag\\
&\qquad \qquad \left.  -\iint (y - x) f(\max\{0, t-c^{-1}|y-x|\}, y, p) \, dp \frac{dy}{|y-x|^3} \right| \notag\\
& \qquad \lesssim \int_{|y|\leq R+\mathcal{P}(t)t} \int_{|p|\leq \mathcal{P}(t)} \int^t_{\max\{0, t-c^{-1}|y-x|\}}
  		 							 |\partial_t f_\infty(s, y, p)| \,ds \, dp \frac{dy}{|y-x|^2}\notag\\
& \qquad \qquad + \int_{|y|\leq R+\mathcal{P}(t)t} \int_{|p|\leq \mathcal{P}(t)} D_F(\max\{0, t-c^{-1}|y-x|\})\, dp \frac{dy}{|y-x|^2}  + c^{-1}. \notag\\
& \qquad \lesssim D_F(t) + c^{-1}. \label{phix_Ux}
\end{align}
To estimate $D_F(t)$, let us define $D_f:= f - f_\infty$. Then using the two Vlasov equations (\ref{NV2}) and (\ref{VP1}),
we obtain
\begin{align}
&\partial_t D_f +  \widehat{p}\cdot\nabla_x D_f
		   - \big[S(\phi)p+\frac{c^2\nabla_x\phi}{\sqcp}\big]\cdot\nabla_p D_f  \label{Df}\\
& \quad = (p-\widehat{p})\cdot\nabla_xf_\infty 
  		  + \big[S(\phi)p+\frac{c^2\nabla_x\phi}{\sqcp} -\nabla_xU\big]\cdot\nabla_p f_\infty + 4 S(\phi) (f_\infty + D_f). \notag
\end{align}
Note that $|p-\widehat{p}| \leq c^{-2}\mathcal{P}_c(t)^3$.
Also note that $\nabla_xf_\infty$, $\nabla_p f_\infty$ and $f_\infty$ are bounded on $\mathbb{R}^6 \times [0, T']$. 
Then with (\ref{phix_Ux}), (\ref{Df}) becomes
\begin{align}
&\Big|\partial_t D_f +  \widehat{p}\cdot\nabla_x D_f
		   - \big[S(\phi)p+\frac{c^2\nabla_x\phi}{\sqcp}\big]\cdot\nabla_p D_f \Big| \notag\\
& \lesssim c^{-1} + |c^2 \nabla_x \phi - \nabla_x U| + c^{-1} |D_f|
 \lesssim D_F(t) + c^{-1}. \label{DF1}
\end{align}
Using the characteristics $(X, P)(s)$ of (\ref{NV2}) with $(X, P)(t) = (x, p)$, compute
$$
\Big|\frac{d}{ds}D_f(s, X(s), P(s)) \Big| \lesssim D_F(s) + c^{-1}.
$$
Note that $D_f(0, X(0), P(0)) = 0.$ 
Therefore integrating (\ref{DF1}) we get
\begin{align}
D_F(t) \lesssim \int^t_0 D_F(s) \,ds + c^{-1}. 
\end{align}
So Gronwall's inequality implies $D_F(t) \lesssim c^{-1}$, which gives (\ref{newlimit2}). The proof of (\ref{newlimit1}) is completed by
(\ref{phi_U_final}) and (\ref{phix_Ux}). This concludes the proof of Theorem \ref{newlim}.

\subsection*{Acknowledgments}
The authors thank Gerhard Rein and Alan~D.~Rendall for comments on the manuscript and suggestions.
S.~C.\ acknowledges the Albert Einstein Institute for the kind hospitality and the European network HYKE for financial support 
(contract HPRN-CT-2002-00282). 


\end{document}